\documentclass[10pt]{article}
\usepackage{amsmath}
\usepackage{amssymb}
\usepackage{graphicx}

\def\apj{ApJ }

\def\mnras{MNRAS }

\begin{document}

\setcounter{figure}{0}
\setcounter{table}{0}
\setcounter{footnote}{0}
\setcounter{equation}{0}

\vspace*{0.5cm}

\noindent {\Large ENSEMBLE PULSAR TIME SCALE}
\vspace*{0.7cm}

\noindent\hspace*{1.5cm} A.E. RODIN$^1$\\
\noindent\hspace*{1.5cm} $^1$ Pushchino Radio Astronomy Observatory of the
Lebedev Physical Institute RAS\\
\noindent\hspace*{1.5cm} Russia, 142290, Moscow region, Pushchino, PRAO\\
\noindent\hspace*{1.5cm} e-mail: rodin@prao.ru\\

\vspace*{0.5cm}

\noindent {\large ABSTRACT.} The algorithm of the ensemble pulsar time scale (PT$_{\rm ens}$) based on
the optimal Wiener filtration method has been proposed. This algorithm allows
the separation of the contributions to the post-fit pulsar timing residuals of
the atomic clock and pulsar itself. Filters were designed with the use of the
cross-spectra of the timing residuals. The method has been 
applied to the timing data of six millisecond pulsars. Direct comparison with the
classical method of the weighted average showed that use of the optimal Wiener filters before averaging allows
noticeably to improve the fractional instability of the ensemble time scale. Application of the proposed method to the most stable millisecond pulsars with the fractional instability $\sigma_z < 10^{-15}$ may improve the fractional instability of PT$_{\rm ens}$ up to the level $\sim 10^{-16}$.

\vspace*{1cm}

\noindent {\large 1. Introduction}

\smallskip

Despite the algorithms of construction of the ensemble time scales are well established and the fractional instability of the TAI and TT(BIPM) scales currently is at level of a few $10^{-16}$ (Petit, 2010), development of new methods and approaches of the ensemble scales construction, including pulsar ones, presents an interest (Rodin, 2008), (Hobbs et al, 2010). In this paper the method of the optimal Wiener filtration is developed for construction of the ensemble pulsar time scale as applied to pulsar timing data. This method is compared with the classical algorithm based on the computation of the weighted average of time scales. From the general reasoning one can expect increasing of the signal estimation accuracy in comparison with the weighted average method since an additional information is used in form of the covariance function or spectrum of the estimated signal. The paper (Rodin, 2008) on the basis of computer simulation confirms the above statement.

In this paper PT$_{\rm ens}$ is considered primary as an independent basement for calculation of variations of the terrestrial time scales rather than an independent realisation of the barycentric time.

Section 2 of the paper contains basic formulae of the optimal Wiener filtration. Section 3 tells about pulsar observations used in this paper. In section 4 the method of optimal filtration is applied to timing data of six millisecond pulsars. The fractional instability $\sigma_z$ is computed for time scales averaged with the different methods.

\vspace*{0.7cm}

\noindent {\large 2. Basic formalae}

\smallskip

Let us assume that we have $M$ time series of the length $n$  ${}^k r(t_i)={}^k r_i(t)\;(i=1,2,\ldots,n,\,k=1,2,\ldots,M)$. In our case ${}^k r_i$ are post-fit timing residuals of time of arrivals (TOAs) of pulsar pulses observed relative to the same time scale and with the same registering equipment. This paper uses so called square root Wiener filter which expressed by the following formula (Terebizh, 1993)
\begin{equation}\label{filter}
{}^k H(\omega)=\sqrt{\frac{S(\omega)}{S(\omega)+{}^k N(\omega)}},\;
k=1,2,\ldots,M,
\end{equation}
where $S(\omega)$ is spectrum of the signal, ${}^k N(\omega)$ is spectrum of the noise of the $k$th pulsar.

In this paper the problem of stochastic signal estimation is solved under condition of lack of apriori information, since the covariance matrix and spectrum of the signal are apriori unknown and estimated from the data itself in the assumption that the clock variations (estimated signal) and variations of the rotational pulsar phase (additive noise) are uncorrelated. Spectrum $S(\omega)$ of the signal were calculated as an average of all cross-spectra by the formula ($k\neq l$)
\begin{equation}\label{spxy}
S(\omega)=\frac{1}{2\pi}\left|^kR(\omega)^lR^{*}(\omega) \right|,\; k,l=1,2,\ldots,M,
\end{equation}
where $(\cdot)^*$ denominates complex conjugation, ${}^k R(\omega)={\cal F}[{}^k r(t)]$ is Fourier transform of the input data.

The optimal filter is used according to the following formula
\begin{equation}
{}^k \hat R(\omega)={}^k H(\omega){}^k R(\omega),\; k=1,2,\ldots,M.
\end{equation}
The filtered signal is obtained with the inverse Fourier transform ${}^k s(t)={\cal F}^{-1}[{}^k \hat R(\omega)]$.

The averaged signal (the ensemble scale) is calculated by the weighted average formula 
\begin{equation}
{\bf s}=\sum_{k=1}^M {}^kw {}^k s(t),\;\sum_k {}^k w=1.
\end{equation}
Weight ${}^kw\sim {}^k{\rm rms}^{-2}$, rms is the root mean square of the $k$th time series.

\vspace*{0.7cm}

\noindent {\large 3. Observations}

\smallskip

The pulsars  PSR J0613-0200, J1640+2224, J1643-1224, J1713+0747, J1939+2134 and J2145-0750 were observed with the fully steerable RT-64 radio telescope of the Kalyazin Radio Astronomy Observatory (KRAO) (Potapov et al., 2003) (Ilyasov et al., 2004), (Ilyasov et al., 2005) (fig.~1). The AS-600 instrumental facility of the Pushchino Radio Astronomy Observatory (Astro Space Center, Lebedev Physical Institute) was used for the registration (Oreshko, 2000). The pulsar pulses were accumulated by a spectrum analyzer in two circular polarizations, 80 channels in each, with a channel frequency band 40 kHz. The observing sessions were conducted, on average, once every two weeks. The total signal integration time in each session was about 2 hours.

The topocentric TOAs were determined by fitting the session-summed pulsar pulse profile into a reference template with a hight signal to noise ratio. We computed the barycentric TOAs, determined the TOA residulas and refined the pulsar timing parameters by minimizing the residuals by the least-squares method using the software Tempo (Taylor, Weisberg, 1989). The DD model (Damour, Deruelle, 1986) was used to refine and compute the pulsar orbital parameteres. The astrometric, spin and orbital parameters of all six pulsars can be found in the paper (Rodin, 2011).

\begin{figure}\label{fig1}
\includegraphics[scale=0.7]{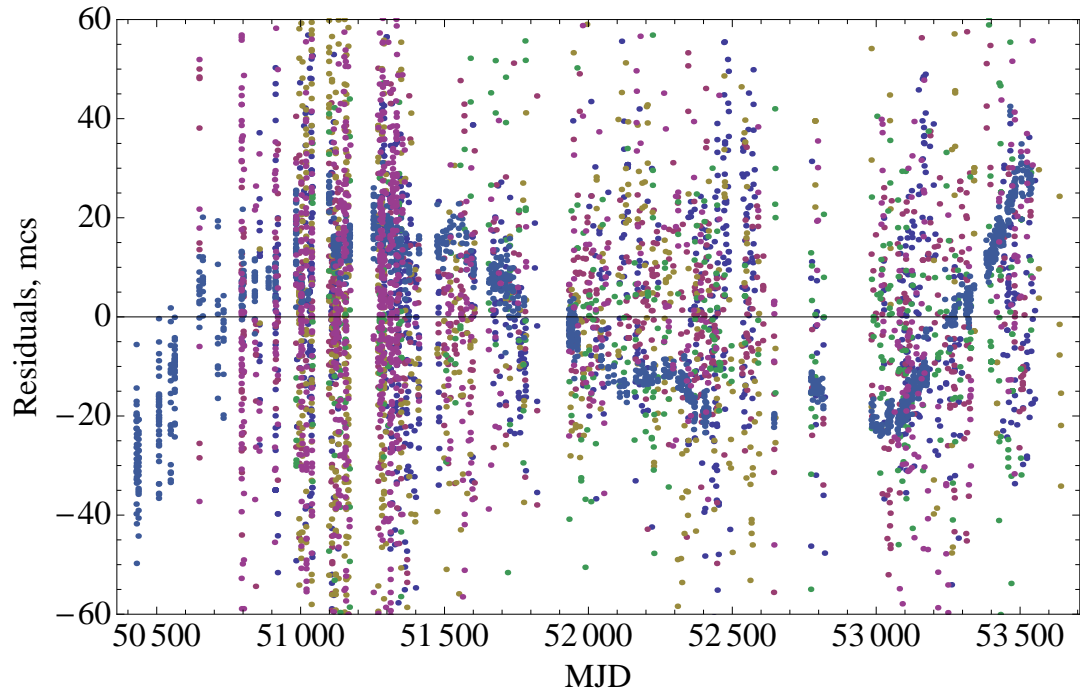}
\caption{Post-fit timing residuals of six millisecond pulsars observed at the Kalyazin Radio Astronomy Observatory.}
\end{figure}

\vspace*{0.7cm}

\noindent {\large 4. Results}

\smallskip

For the purpose of unification all pulsar data were binned and averaged at the interval 30 days. Gaps were filled with the linear interpolation of adjacent values. The common part of the data in the interval MJD=51000--53490 were used. Since all observations were carried out with the same registering system, no data matching was applied.

Fig.~2a shows the binned pulsar residuals and their weighted average (solid line). Fig.~2b shows the same data passed through the optimal filter (\ref{filter}) and their weighted average.

\begin{figure}\label{fig2}
\begin{minipage}{0.5\linewidth}
\vbox{\hbox{\hspace*{4.4cm}(a)}\includegraphics[scale=0.75]{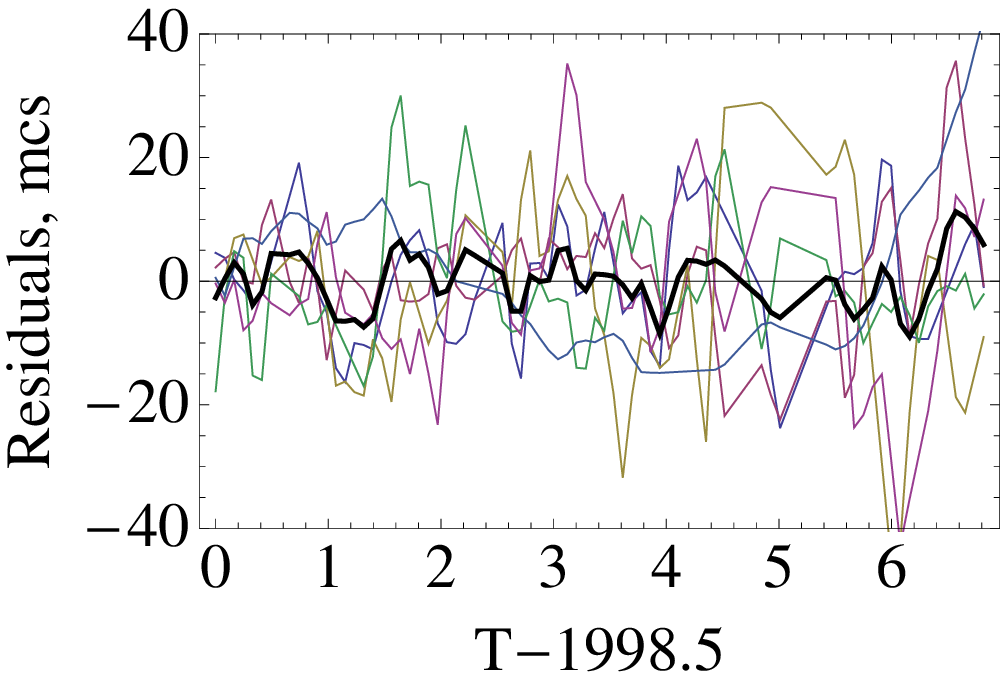}}
\end{minipage}
\hfill
\begin{minipage}{0.5\linewidth}
\vbox{\hbox{\hspace*{4.4cm}(b)}\includegraphics[scale=0.75]{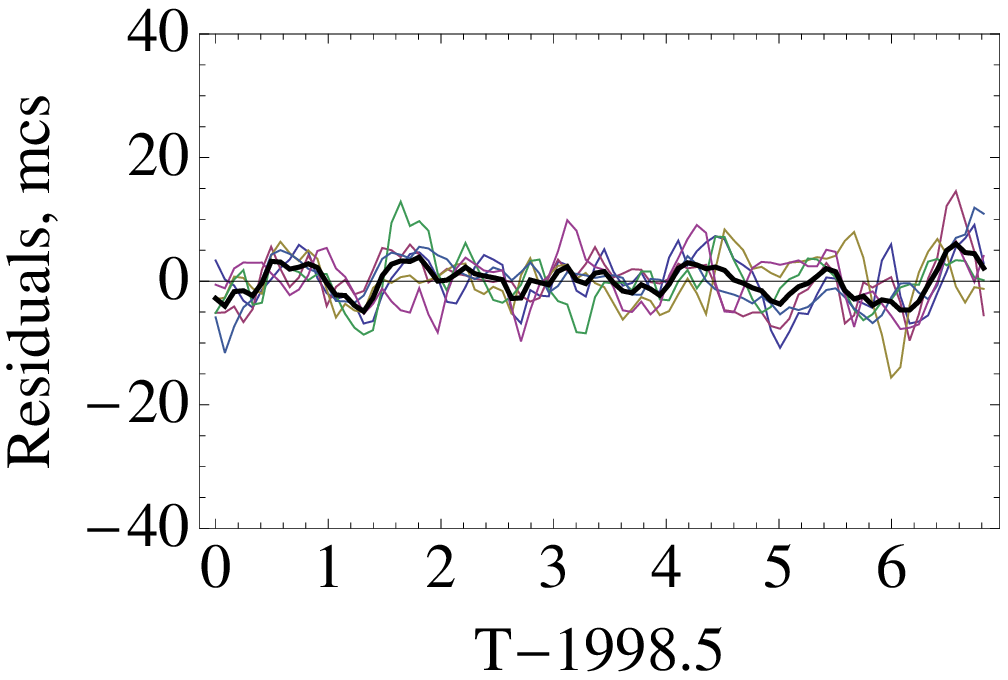}}
\end{minipage}
\caption{The binned post-fit timing residuals of six millisecond pulsars observed at the Kalyazin Radio Astronomy Observatory before (a) and after (b) applying Wiener optimal filter. Solid line indicates weighted average.
}
\end{figure}

The stability of a time scale is characterised by so-called Allan
variance numerically expressed as a second-order difference of the
clock phase variations. Since timing analysis usually includes
determination of the pulsar spin parameters up to at least the first
derivative of the rotational frequency, it is equivalent to excluding
the second order derivative from pulsar TOA residuals and therefore
there is no sense in the Allan variance. For this reason, for
calculation of the fractional instability of a pulsar as a clock,
another statistic $\sigma_z$ has been proposed (Taylor, 1991).

Fig.~3 shows the fractional instability of the ensemble pulsar time scale constructed on the basis of six millisecond pulsars observed at KRAO. From the fig.~3 one can see that the optimal filter applied to data before weighted average improves the fractional instability almost two times along all time interval $\tau$. On the basis of visual analysis of the fig.~2 one can conclude that the common signal presented in all pulsar data and caused by behavior of the registering equipment and by variations of the local frequency standard is better determined in data passed through the filter.

\begin{figure}\label{fig3}
\includegraphics[width=8cm]{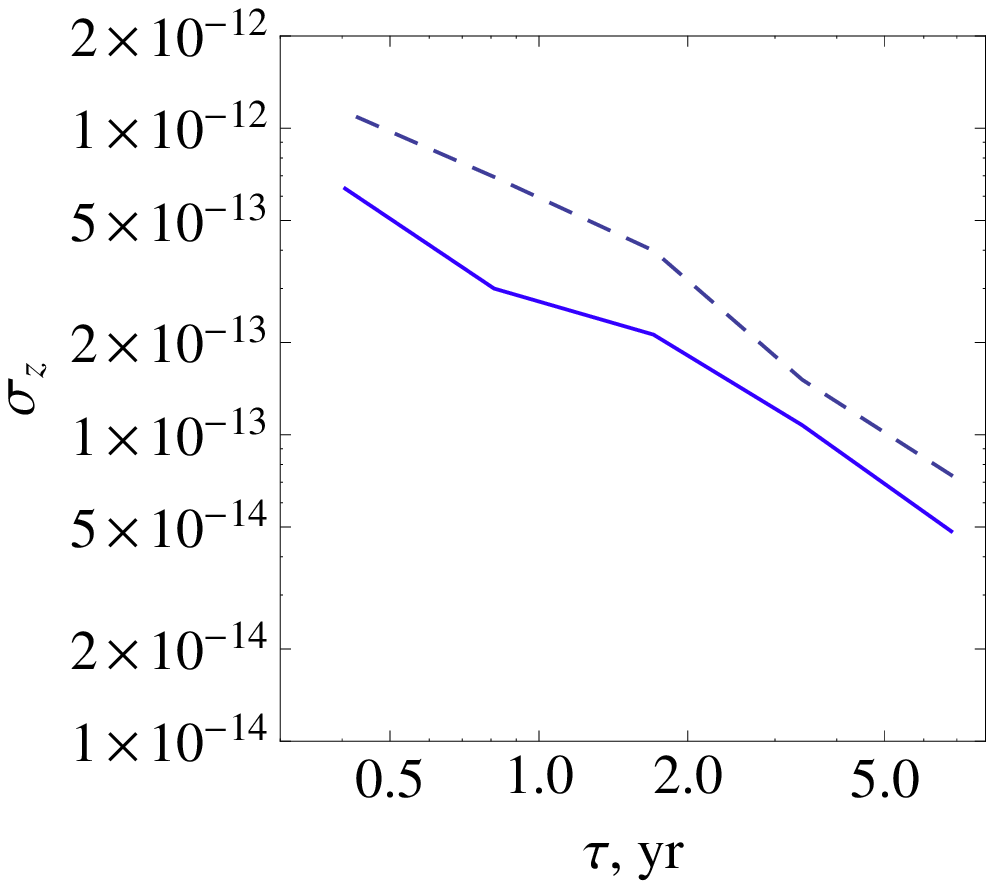}
\caption{Fractional instability $\sigma_z$ of the ensemble pulsar time scale in dependence on the observation interval $\tau$. Dashed line indicates weighted average method, solid line indicates optimal filters method.}
\end{figure}

At the present time at least five pulsars display the fractional instability at the level $\leq 10^{-15}$ (Verbiest et al, 2009). One can expect that application to them of the optimal filter method will allow to obtain the fractional instability of the ensemble pulsar time scale at the level of a few units of $10^{-16}$ , i.e. comparable with the instability of the best terrestrial frequency standards.

\vspace*{0.7cm}

\noindent {\large 5. REFERENCES}
%
%
%
%
%
{

\leftskip=5mm
\parindent=-5mm

\smallskip

Damour D., Deruelle N, 1986, ``General relativistic celestial mechanics of binary systems. II. The post-Newtonian timing formula'', Ann. Inst. Henri Poincare, Physicue theorique, {44}, 263 (1986).

Hobbs, G.; Coles, W.; Manchester, R.; Chen, D., 2010, ``Developing a pulsar-based timescale'', this issue.

Ilyasov Yu., Oreshko V., Potapov V., Rodin A., 2004, ``Timing of Binary Pulsars at Kalyazin, Russia'', IAU Symposium no.218, (Ed. F.~Camilo and B.~M.~Gaensler, ASP Series, San Francisco, 2004), p.~433.

Ilyasov Yu., Imae M., Hanado Y., Oreshko V., Potapov V., Rodin A., Sekido M., 2005, ``Two-frequency timing of the pulsar B1937+21 in Kalyazin and Kashima in 1997-2002'', Astron Lett., {31}, 33.

Oreshko.V.V., 2000, ``Pulsar timing instrumental errors. AC-600/1600 facility''. Proceedings of the Lebedev Physical Institute.,(Ed. O.Krokhin and N.Kardashev, Moscow, 2000) v. 229, p. 110 (in Russian).

Petit G., 2010, ``Atomic time scales TAI and TT(BIPM): present performances and prospects'', Highlights of Astr., 15, p. 220.

Potapov, V. A.; Ilyasov, Yu. P.; Oreshko, V. V.; Rodin, A. E., 2003, ``Timing Results for the Binary Millisecond Pulsar J1640+2224'', Astronomy Letters, 29, p. 241.

Rodin A.E., 2008, ``Optimal filters for the construction of the ensemble pulsar time'', \mnras, {387}, 1583.

Rodin A.E., 2011, ``Detection of the gravitational waves by observations of several pulsars'', Astr.Lett., v.88, No.2, p.1.

Taylor J.~H. and Weisberg J.~M., 1989, ``Further experimental tests of relativistic gravity using the binary pulsar PSR 1913 + 16'', \apj {345}, 434.

Taylor J.H., 1991, ``Millisecond pulsars - Nature's most stable clocks'', Proc.IEEE {79}, 1054.

Terebizh V.Yu., 1992, ``Analysis of time series in astrophisics'', Moscow: Nauka (in Russian).

Verbiest, J. P. W.; Bailes, M.; Coles, W. A.; Hobbs, G. B.; van Straten, W.; Champion, D. J.; Jenet, F. A.; Manchester, R. N.; Bhat, N. D. R.; Sarkissian, J. M.; Yardley, D.; Burke-Spolaor, S.; Hotan, A. W.; You, X. P., 2009, ``Timing stability of millisecond pulsars and prospects for gravitational-wave detection'', \mnras, 400, Issue 2, p. 951.

}

\end{document}